\documentclass[aps,pre,tightenlines,notitlepage,nofootinbib,superscriptaddress,noshowkeys,11pt]{revtex4-1}

\usepackage{amsmath,subeqnarray}
\usepackage{amssymb,amsfonts}
\usepackage{graphicx}
\usepackage{bm}
\usepackage{subcaption}
\usepackage{color}
\usepackage[colorlinks=true]{hyperref}
\usepackage[normalem]{ulem}

\graphicspath{{figs/}}

\newcommand{\mathnotation}[2]{\newcommand{#1}{\ensuremath{#2}}}

\newcommand{\Order}[1]{\mathrm{O}(#1)}      

\renewcommand{\l}{\left}
\renewcommand{\r}{\right}
\renewcommand{\time}{t}
\mathnotation{\pd}{\partial}
\mathnotation{\ee}{\mathrm{e}}              
\mathnotation{\imi}{\mathrm{i}}             
\mathnotation{\dint}{\,{\mathrm{d}}}        

\mathnotation{\grad}{\nabla}
\renewcommand{\div}{\grad\cdot}
\mathnotation{\lapl}{\grad^2}

\mathnotation{\Diff}{D}                     
\mathnotation{\RR}{a}                       
\mathnotation{\Uc}{U}                       
\mathnotation{\uc}{u}                       
\mathnotation{\uv}{\bm{\uc}}                
\mathnotation{\pres}{p}                     

\mathnotation{\rr}{r}                       
\mathnotation{\xc}{x}                       
\mathnotation{\yc}{y}                       
\mathnotation{\zc}{z}                       
\mathnotation{\rv}{\bm{\rr}}                
\mathnotation{\thetauv}{\hat{\bm{\theta}}}  
\mathnotation{\xuv}{\skew{4}\hat{\bm{\xc}}} 
\mathnotation{\ruv}{\hat{\bm{r}}}           

\mathnotation{\C}{C}                        
\mathnotation{\fl}{f}                       
\mathnotation{\F}{F}                        
\mathnotation{\s}{s}                        
\mathnotation{\fff}{\varphi}                
\mathnotation{\T}{T}                        
\mathnotation{\LP}{P}                       
\mathnotation{\g}{g}                        
\mathnotation{\gh}{\widehat{g}}             
\mathnotation{\Cc}{\mathcal{C}}             
\mathnotation{\dr}{\Delta\rr}               
\mathnotation{\dth}{\Delta\theta}           
\mathnotation{\dt}{\Delta t}                
\def\v{\vspace{1cm}}
\mathnotation{\Pe}{\mathrm{Pe}}

\begin{document}

\title{Can phoretic particles swim in two dimensions?}

\author{David Sondak}
\email{dsondak@ices.utexas.edu}
\author{Cory Hawley}
\author{Siyu Heng}
\author{Rebecca Vinsonhaler}
\affiliation{Department of Mathematics, University of Wisconsin --
  Madison, Madison, WI, USA}
\author{Eric Lauga}
\email{e.lauga@damtp.cam.ac.uk}
\affiliation{Department of Applied Mathematics and Theoretical Physics,
  University of Cambridge, Cambridge, UK}
\author{Jean-Luc Thiffeault}
\email{jeanluc@math.wisc.edu}
\affiliation{Department of Mathematics, University of Wisconsin --
  Madison, Madison, WI, USA}

\date{\today}

\begin{abstract}
Artificial phoretic particles swim using self-generated gradients in chemical species (self-diffusiophoresis) or charges and currents (self-electrophoresis). These particles can be used to study the physics of collective motion in active matter and might have promising applications in bioengineering. In the case of self-diffusiophoresis, the classical physical model relies on a steady solution of the diffusion equation, from which chemical gradients, phoretic flows and ultimately the swimming velocity, may be  derived.   Motivated by disk-shaped particles in thin films and under confinement, we examine  the extension to two dimensions. Because the  two-dimensional diffusion equation lacks a steady state with the correct    boundary conditions, Laplace transforms must be used to study the long-time behavior of the problem and  determine the swimming velocity.  For fixed chemical fluxes on the particle surface, we find that the swimming velocity  ultimately always decays logarithmically in time.   In the case of finite P\'eclet numbers, we  solve the full advection-diffusion   equation numerically and show that this decay can be avoided by the particle moving to  regions of unconsumed reactant.  Finite advection thus regularizes the two-dimensional phoretic problem.

\v

\end{abstract}

\keywords{Self-propelled janus particles; Low Reynolds number swimmers; Microswimmers; Diffusiophoresis; Locomotion}

\maketitle

\section{Introduction}

One of the most exciting recent developments in soft matter research has been the flurry of work on the physics of active matter \cite{ramaswamy10}. Originally focusing on problems in  biological physics, specifically the study of self-propelled microorganisms \cite{lighthill76,brennen77,lauga2009}, active matter encompasses now a variety of interacting, driven systems including colloids, gels, granular flows, and populations   \cite{marchetti13}.

One category of active matter comprises artificial, small-scale swimmers \cite{Wang2009,nelson2010}. In order to drive an artificial swimmer, one would typically either (a) rely on external energy sources, e.g.~two-\cite{dreyfus05} or three-dimensional \cite{Zhang2010b,Pak2011} magnetic fields, or acoustic fields \cite{wang2012,nadal2014}; or (b)  take advantage of a local energy source in the form of chemical reactions. While recent work has shown that chemistry is able to propel large active droplets via Marangoni instabilities   \cite{thutupalli2011,izri2014}, the majority of the work on chemical propulsion takes advantage of self-electrophoresis and self-diffusiophoresis mechanisms \cite{anderson1989}.
Using phoretic flows to induce swimming was first demonstrated experimentally with Janus platinum-gold colloidal rods undergoing directed motion in aqueous solutions of hydrogen peroxide whose reduction to water was catalyzed by the platinum side of the rods  \cite{paxton2004}. Since this seminal study, much has been devoted to further understanding catalytic swimming both experimentally~\cite{Howse2007,Ebbens2011,ebbens2012} and theoretically~\cite{golestanian2005,golestanian2007,cordova2008,julicher2009,brady2011,sabass2012}.  Recent work showed how to bypass the use of  catalyst gradients using solely geometry~\cite{shklyaev2014,michelin2015}, how to assemble phoretic crystals~\cite{palacci2013}, the subtle role of electrokinetic effects~\cite{brown2014,ebbens2014}, and the impact of solute advection~\cite{michelin2013c,michelin2014}.

In the simplest  continuum model of a phoretic swimmer, the solute (or reactant) satisfies a diffusion equation with a fixed flux on a portion of the surface of the swimmer, modeling its creation or absorption at a fixed rate. This simple chemical approach implicitly assumes that the reactant never gets  depleted.  The swimming velocity is then obtained by integrating the flow induced locally tangentially to the surface of the swimmer by the  chemical gradients \cite{anderson1989} in a way that satisfies the force-free condition at all times  \cite{stone96}.

This simple modeling approach works very well in three dimensions (3D), but there is a mathematical problem with attempting to the do the same in two dimensions (2D). Indeed, while the solution to the diffusion equation with a net flux admits a steady-state solution decaying at infinity in 3D, it does not in 2D because the solution is logarithmically growing. This feature is actually intrinsically tied to the recurrence properties of random walks or Brownian motion as a function of dimension.  One mathematical way out of this conundrum is to require the surface to not be a net source (or sink) of either solute or reactant~\cite{crowdy2013}, but this is of course a restrictive  assumption.

From an applied standpoint, it could be however possible to  carry out phoretic experiments closely mimicking two dimensions, for example with phoretic disks in a Hele--Shaw cell or in a freely-suspended thin film.  Phoretic swimming in two dimensions thus raises some interesting questions.  Can 2D Janus particles undergo steady swimming or will their swimming speed always be time-dependent?  Can solute advection regularize this apparent 2D peculiarity?  Is chemical depletion and a modification of the boundary conditions the key ingredient?

This paper attempts to address these  questions.  We begin, in Section~\ref{sec:bg} by reviewing the theory of locomotion of a three-dimensional Janus particle.  In Section~\ref{sec:noadv} we perform an asymptotic analysis on the two-dimensional diffusion equation for a circular Janus particle with surface flux boundary conditions, using the Laplace transform as our primary tool.  We then determine the asymptotic swimming speed of the particle assuming infinite solute concentration (Section~\ref{sec:BC1}) and finite solute concentration (Section~\ref{sec:BC2}).  Following this analysis, we consider solute advection numerically in Section~\ref{sec:full}.  We review our numerical approach in Sections~\ref{sec:vel}--\ref{sec:numerics} before presenting results in Section~\ref{sec:results}.  We conclude by summarizing our findings in Section~\ref{sec:conclusions}.

\section{Phoretic swimming in  three dimensions}
\label{sec:bg}

We start by  reviewing the classical continuum model for locomotion of three-dimensional phoretic particles. We ignore  electrophoretic effects and focus on the case of neutral solutes for which locomotion is driven by self-diffusiophoresis~\cite{golestanian2005,golestanian2007}.  Let~$\varphi(\rv,\time)$ denote the concentration of  a reactant with diffusivity~$D$ outside an isolated spherical particle of radius~$\RR$ in an infinite fluid.  For a spherical particle with an axially-symmetric coating and in the absence of advection by the flow, the concentration~$\varphi(\rr,\theta,\time)$ obeys the diffusion equation in spherical coordinates
\begin{equation}
	\frac{\pd\varphi}{\pd\time} = \Diff\lapl\varphi
	= \Diff\l[\frac{1}{\rr^2}
	\frac{\pd}{\pd\rr}\l(\rr^2\frac{\pd\varphi}{\pd\rr}\r)
	+
	\frac{1}{\rr^2\sin\theta}\,
	\frac{\pd}{\pd\theta}\l(\sin\theta\,\frac{\pd\varphi}{\pd\theta}\r)
	\r],
	\qquad \rr \ge \RR.
	\label{eq:heat3}
\end{equation}
The simplest description of the chemical boundary condition at the surface of the particle is to assume that the flux of reactant is prescribed, constant in time and spatially-dependent as
\begin{subequations}
  \begin{align}
    -\ruv\cdot\grad\varphi(\RR,\theta,\time) &= \RR^{-1}\fl(\theta);
    \label{eq:bcflux} \\
    \lim_{r\to\infty}\varphi(r,\theta,\time) &= 0
    \label{eq:bcinfty}
  \end{align}
  \label{eq:bc3d}%
\end{subequations}
for~$0\leq\theta \leq \pi$.
Note that we have absorbed a factor~$\Diff/\RR$ in our definition of~$\fl$, so the flux has the same units as~$\varphi$.  Wherever~$\fl(\theta)<0$, the particle is absorbing the reactant~$\varphi$.  The zero boundary condition at infinity in~\eqref{eq:bcinfty} is a mathematical convenience; in an experiment, there is a concentration~$\C>0$ of some reactant at  infinity, and~$\varphi < 0$ measures the deficit in that reactant.  Note that axial symmetry implies that the third spherical coordinate ($\phi$) does not enter the problem, so all fields are axisymmetric and swimming is restricted to happen on the symmetry axis of the sphere ($z$).

The steady-state solution to~\eqref{eq:heat3} with boundary conditions~\eqref{eq:bc3d} is classically given by
\begin{equation}
    \varphi(\rr,\theta) = \sum_{m=0}^\infty
    \F_m\,(\RR/\rr)^{m+1}\,\LP_m(\cos\theta),
\end{equation}
where~$\LP_m(x)$ are the Legendre polynomials with the usual
normalization~$\LP_m(1)=1$, and where
\begin{equation}
  \F_m = \frac{2m+1}{2(m+1)}
  \int_{-1}^1\fl(\theta)\,\LP_m(\cos\theta)\dint(\cos\theta).
\end{equation}
Note that~$\F_0$ is the mean of~$\fl(\theta)$ over the spherical particle.

The local fluid velocity at the surface of the particle  is in the tangential direction and is proportional  to the reactant concentration via
\begin{equation}
  \uv(\RR,\theta) = \mu\,\pd_\theta\varphi\l(\RR,\theta\r)\,\thetauv
  \label{eq:uvtheta},
\end{equation}
where $\mu$ is the surface phoretic mobility, which could have either sign depending on the details of the short-range interactions between the reactant and the surface \cite{anderson1989}.  Since the particle is circular, its swimming velocity  is obtained by averaging~$-\uv$ along its surface (see Refs.~\cite{stone96,Elfring2015} for the extension to more complex shapes).  The averaged component of velocity perpendicular to the~$\zc$ axis vanishes by symmetry so we obtain ${\bf U}=\Uc{\bf z}$ with
\begin{equation}
  \Uc = -\tfrac23\mu\F_1.
  \label{eq:swimvel3D}
\end{equation}
Hence, in the absence of advection of the reactant by the flow field (zero-P\'eclet number limit), a spherical phoretic particle swims with  the steady velocity given by~\eqref{eq:swimvel3D}.  In the next sections, we address how this is modified in two dimensions.

\section{Phoretic swimming in  two dimensions: No advection}
\label{sec:noadv}
In many hydrodynamics problems, the two-dimensional analysis is carried out before the extension to three dimensions. In the case of phoretic swimming, we in fact go the other way and address the subtle two-dimensional case after the easier three-dimensional analysis.

\subsection{Parameters and dimensionless groups}

As in the three-dimensional problem, there are three relevant dimensional scales: length, time, and units of the concentration deficit, $\varphi$.  There are five physical parameters: the particle radius, $\RR$, the reactant diffusivity, $\Diff$, the background concentration, $\C$, the characteristic reactant surface flux, $\F$, and the phoretic mobility from
Eq.~\eqref{eq:uvtheta}, $\mu$.  Thus, there are two dimensionless groups, which we take to be~$\C/(-\F)$ and~$(-\mu\F)/\Diff$.  We shall keep~$\F<0$ throughout, indicating a sink of reactant.  In all the numerical examples, we will take~$\RR=\Diff=-\F=1$, so that~$\C$ and~$\mu$ stand in for the two dimensionless numbers and are thus the only two parameters to be varied.  In these units, the velocity~$\Uc$ is equal to the P\'eclet number, $\Pe=\Uc\RR/\Diff$.

\subsection{General solution using Laplace transform}
\label{sec:Laplace2D}

We consider a circular Janus particle in two spatial dimensions.  The concentration~$\varphi(\rr,\theta,\time)$ obeys a diffusion equation in cylindrical coordinates,
\begin{equation}
  \frac{\pd\varphi}{\pd\time} = \Diff\lapl\varphi
  = \Diff\l[\frac{1}{\rr}
    \frac{\pd}{\pd\rr}\l(\rr\frac{\pd\varphi}{\pd\rr}\r)
    +
    \frac{1}{\rr^2}\,\frac{\pd^2\varphi}{\pd\theta^2}
    \r],
  \qquad \rr \ge \RR, \quad -\pi \leq \theta \leq \pi.
  \label{eq:heat2}
\end{equation}
We will consider two different boundary conditions in Sections~\ref{sec:BC1} and~\ref{sec:BC2}.  In either case, unlike the three-dimensional case, Eq.~\eqref{eq:heat2} with accompanying boundary conditions does not  have a steady solution decaying to infinity if it is a net sink (or source) of reactant.
Hence, in order to understand the ultimate fate of two-dimensional particles we must solve the full time-dependent problem.  Here, we will use Laplace transforms because we are mostly interested in long-time asymptotics.

The Laplace transform of~\eqref{eq:heat2} gives
\begin{equation}
\s\, \tilde\varphi - \varphi(\rr,\theta,0)
=
\Diff\l(\frac{1}{\rr}
\frac{\pd}{\pd\rr}\l(\rr\frac{\pd\tilde\varphi}{\pd\rr}\r)
+
\frac{1}{\rr^2}\,\frac{\pd^2\tilde\varphi}{\pd\theta^2}
\r),
\label{eq:heat2LT}
\end{equation}
where~$\tilde\varphi(\rr,\theta,\s)$ is the Laplace transform of~$\varphi(\rr,\theta,\time)$.  For simplicity  we take~$\varphi(\rr,\theta,0) = 0$.  Equation~\eqref{eq:heat2LT} has a solution in terms of modified Bessel functions of the second kind,
\begin{equation}
\tilde\varphi(\rr,\theta,\s)
=
\sum_{m=0}^\infty K_m(\rr\sqrt{\s/\Diff})\,
(A_m\cos m\theta + B_m\sin m\theta),
\label{eq:Ksol}
\end{equation}
where we applied the boundary condition at infinity~\eqref{eq:bcinfty}.  In the following subsections we apply two types of boundary conditions at~$\rr=\RR$ to \eqref{eq:Ksol} and examine the resulting solutions.

\subsection{Fixed flux boundary condition}
\label{sec:BC1}
In this section, we first consider the case of the inhomogeneous Neumann boundary condition,~\eqref{eq:bcflux}, where the reactant is assumed to be consumed at a fixed flux.
Using the Laplace-transformed solution of the two-dimensional diffusion equation, we   perform an asymptotic analysis on the solution to determine the large-time behavior of the particle.

The Neumann boundary condition at~$\rr=\RR$, adapted to two dimensions from~\eqref{eq:bcflux}, has Laplace transform
\begin{align}
-\ruv\cdot\grad\tilde\varphi(\RR,\theta,\s) = \fl(\theta)/\l(\RR\s\r)
\end{align}
which we apply to~\eqref{eq:Ksol} to obtain
\begin{equation}
\sum_{m=0}^\infty \sqrt{\s/\Diff}\,K_m'(\RR\sqrt{\s/\Diff})\,
(A_m\cos m\theta + B_m\sin m\theta) = -\fl(\theta)/\RR\s
\label{eq:bcapply},
\end{equation}
where~$K_m'(\cdot)$ is the derivative of $K_m(x)$ with respect to $x$.  To parallel the three-dimensional case, we assume that~$\fl(\theta)$ is even, which corresponds to a symmetric particle.  Then we directly obtain $B_m=0$ and
\begin{equation}
A_m
= -\frac{\sqrt{\Diff/\RR^2\s^3}}{K_m'(\RR\sqrt{\s/\Diff})}\,\F_m\,,
\end{equation}
with
\begin{equation}
    \F_m = \left\{
    \begin{array}{lr}
      \displaystyle \frac{1}{2\pi}\int_{-\pi}^{\pi} \fl(\theta)\dint\theta,
&\quad m = 0; \\[1.25em]
      \displaystyle \frac{1}{\pi}\int_{-\pi}^{\pi}\fl(\theta)\cos
m\theta\dint\theta, &\quad m > 0.
    \end{array}
    \right.
    \label{eq:flcoeff}
\end{equation}
In the present work we will take half of our particle ($-\pi/2 \leq \theta \leq \pi/2$) to be coated with the reactant as a simple test case (two-dimensional Janus particle).  The surface flux is
\begin{equation}
  \fl(\theta) = \left\{
  \begin{array}{lr}
    \F, \quad \lvert\theta\rvert \le \pi/2; \\[0.5em]
    0,  \quad \text{otherwise} .
  \end{array}
  \right.
  \label{eq:surf_flux}
\end{equation}
In that case,
\begin{equation}
\F_0 = \F/2,\quad
\F_{2m} = 0,\quad
\F_{2m-1} = \frac{2}{\pi}\,\frac{(-1)^{m-1}}{2m-1}\,\F, \quad m > 0.
\label{eq:Fhat}
\end{equation}
Substituting for~$A_m$ in~\eqref{eq:Ksol}, we finally get
\begin{equation}
\tilde\varphi(\rr,\theta,\s)
=
-\sqrt{\Diff/\RR^2\s^3}\,
\sum_{m=0}^\infty \F_m\,
\frac{K_m(\rr\sqrt{\s/\Diff})}{K_m'(\RR\sqrt{\s/\Diff})}\,
\cos m\theta.
\label{eq:Ksolbc}
\end{equation}

Instead of inverting the Laplace transform~\eqref{eq:Ksolbc} in order to recover~$\varphi(\rr,\theta,\time)$ to take the $t\to\infty$ limit, we will compute the long-time asymptotics directly by evaluating  the small-$\s$ behavior of~\eqref{eq:Ksolbc}.

To leading order as~$\s \rightarrow 0$, the~$m=0$ term in~\eqref{eq:Ksolbc} is
\begin{equation}
  \tilde\varphi_0(\rr,\theta,\s)
  \sim
  -\F_0\,\s^{-1}(\log(\rr\sqrt{\s/4\Diff}) + \gamma),
  \qquad \s \rightarrow 0,
\end{equation}
where~$\gamma$ is Euler's constant.  This is of the form
\begin{equation}
  \tilde\varphi_0(\rr,\theta,\s) \sim s^{-\rho}\,L(s^{-1})
\end{equation}
where $\rho = 1$ and
\begin{equation}
  L(\xi) = -\F_0\,(\log(\sqrt{\rr^2/4\Diff \xi}) + \gamma)\,.
  \label{eq:Lx}
\end{equation}
A function~$L(\xi)$ is called \emph{slowly-varying at infinity} if~$\lim_{t \rightarrow \infty} L(t \xi)/L(t) \rightarrow 1$ for every fixed~$\xi$.  The function~$L(\xi)$ in~\eqref{eq:Lx} is indeed slowly-varying at infinity, so a Tauberian theorem~\cite[p.~445]{FellerII} gives a formula for the asymptotic antiderivative as
\begin{equation}
  \int\varphi_0(\rr,\theta,\time)\dint\time
  \sim \frac{1}{\Gamma(\rho+1)}\,\time^\rho\,L(\time)
  =
  -\F_0\,\time\,(\log(\sqrt{\rr^2/4\Diff\time}) + \gamma),
  \qquad \time \rightarrow \infty.
\end{equation}
After taking a derivative we obtain the behavior at long times as
\begin{equation}
  \varphi_0(\rr,\theta,\time)
  \sim
  \tfrac12\F_0\,(\log(4\Diff\time/\rr^2) - 2\gamma + 1),
  \qquad \time \rightarrow \infty.
  \label{eq:phi0}
\end{equation}
Note that this analysis is valid for fixed~$\rr$, so it does not satisfy the boundary condition at infinity.  However, we are interested in the behavior near the particle, which corresponds to moderate~$\rr \ge \RR$.  The term~\eqref{eq:phi0} is perfectly well-behaved at~$\rr=\RR$, but it does not contribute to the swimming velocity, since it does not vary with~$\theta$.

For small $s$, the terms in~\eqref{eq:Ksolbc} with $m>0$ are,
\begin{equation}
  \tilde\varphi_m(\rr,\theta,\s) \sim
  \frac{\F_m}{m\s}\,\l(\frac{\RR}{\rr}\r)^m\cos m\theta,
  \qquad
  m>0, \quad \s \rightarrow 0,
\end{equation}
with inverse Laplace transform
\begin{equation}
  \fff_m(\rr,\theta) \sim
  \frac{\F_m}{m}\,\l(\frac{\RR}{\rr}\r)^m\cos m\theta,
  \qquad m > 0.
\end{equation}

The concentration at the surface of the
particle for large~$\time$ is therefore given by
\begin{equation}
  \varphi(\RR,\theta,\time) \sim
  \tfrac12\F_0\left[(\log(4\Diff\time/\RR^2) - 2\gamma + 1\right]
  + \sum_{m=1}^\infty
  \frac{\F_m}{m}\,\cos m\theta,
  \qquad \time \rightarrow \infty.
  \label{eq:phisurf}
\end{equation}
Due to the zero flux boundary conditions~\eqref{eq:bcflux} we must have~$\varphi \ge -\C$, where~$\C$ is the background  reactant concentration, otherwise we will have run out of reactant.
The concentration is lowest at the surface of the particle, and the blowup of the logarithm with time in~\eqref{eq:phisurf} makes this concentration become ever more negative (recall that $F_0 < 0$).  At large times, clearly the  time-independent terms in~\eqref{eq:phisurf} can be neglected compared to the logarithmic term.  We can therefore find a time~$\T$ after which the solute is depleted,  given by
\begin{equation}
  \T = \frac{\RR^2}{4\Diff}\,\exp\l(2\l\lvert\frac{\C}{\F_0}\r\rvert\r).
  \label{eq:T}
\end{equation}
 After this time locomotion should stop.  We note that the exponential dependence should make it possible for this term to be enormous, simply by increasing~$\lvert\C/\F_0\rvert$.

 Let us assume that we are in an intermediate time regime where
\begin{equation}
  1 \ll \frac{4\Diff\time}{\RR^2} \ll
  \exp\l(2\l\lvert\frac{\C}{\F_0}\r\rvert\r),
  \label{eq:trange}
\end{equation}
which allows us to benefit from the large-$\time$ approximation while at the same time ensuring that we do not run out of reactant. In the time range in~\eqref{eq:trange}, we can  proceed with finding the swimming velocity of the particle.  The fluid velocity at the surface is given by~\eqref{eq:uvtheta} while the  swimming velocity of the particle is obtained by averaging~$-\uv$ along its surface.  We note that the local diffusio-osmotic result of~\eqref{eq:uvtheta} does not depend on the dimensionality of the system and is  valid in two dimensions.  The  thickness of the diffuse layer is assumed to be much smaller than the radius of curvature of the particle which allows a matching  between the near field and far field equations for the chemical field, leading to the effective Derjaguin slip velocity of Eq.~\eqref{eq:uvtheta} \cite{anderson1989}.

The averaged~$\yc$ component of velocity vanishes by symmetry.  The averaged~$\xc$ component gives \cite{Elfring2015} 
(with~$\thetauv\cdot\xuv = -\sin\theta$)
\begin{equation}
  \Uc = -\frac{\mu}{2\pi}\int_{-\pi}^{\pi}\pd_{\theta}
        \varphi(\RR, \theta, \time)\l(-\sin \theta\r)\dint\theta
        = -\tfrac12\mu\F_1\,.
  \label{eq:swimvel2D}
\end{equation}
Comparing this result to the 3D case~\eqref{eq:swimvel3D} we observe that the prefactor in front of $\F_{1}$ is smaller for the 2D case.  With the surface flux $\fl(\theta)$ given by~\eqref{eq:surf_flux} we find
\begin{align}
  U = -\mu \F/\pi. \label{eq:Uswim2D_0}
\end{align}
Hence, with half the surface of the particle coated, at long times but not so long that the  reactant is depleted, the particle will swim with the approximately constant steady velocity given by~\eqref{eq:Uswim2D_0}.

\subsection{Concentration-limited flux boundary condition}
\label{sec:BC2}

The criterion~\eqref{eq:trange} tells us that ultimately the equation becomes invalid and  the reactant is  depleted in the vicinity of the swimmer.  It is a simple matter to
modify the boundary condition to reflect this; instead of~\eqref{eq:bcflux} we write
\begin{equation}
  -\RR\,\ruv\cdot\grad\varphi(\RR,\theta,\time)
  = \fl(\theta)(1 + \C^{-1}\varphi(\RR,\theta,\time)).
  \label{eq:bcmod}
\end{equation}
This modification of~\eqref{eq:bcflux} acknowledges that as~$\varphi(\RR,\theta,\time)$ becomes more negative at the surface, the reaction rate must go to zero when~$\varphi(\RR,\theta,\time)=-\C$.  The modified boundary condition~\eqref{eq:bcmod} is related to the classical Michaelis--Menten description of the stationary state of first-order reaction kinetics~\cite{ebbens2012, saha2014clusters}.
Unfortunately, this is more difficult to solve by separation of variables, because the boundary conditions have  nonconstant coefficients.  The Laplace-transformed boundary condition~\eqref{eq:bcmod} gives
\begin{equation}
  \sum_{m=0}^\infty \mathcal{K}_m\,\mathcal{A}_m\cos m\theta =
  -\fl(\theta)\biggl(\s^{-1} + \C^{-1}
    \sum_{m=0}^\infty \mathcal{A}_m\cos m\theta\biggr)
  \label{eq:bigstuff2}
\end{equation}
where
\begin{equation}
  \mathcal{A}_m = K_m(\RR\sqrt{\s/\Diff})\,A_m
\end{equation}
and
\begin{equation}
  \mathcal{K}_m = \RR\sqrt{\s/\Diff}\,K_m'(\RR\sqrt{\s/\Diff})
  /K_m(\RR\sqrt{\s/\Diff}).
\end{equation}
Note that again we have assumed~$\fl(\theta)$ is even (symmetric particle) so that~$B_m=0$.  We can rewrite~\eqref{eq:bigstuff2} as an infinite system
\begin{equation}
\sum_{m=0}^\infty \mathbb{M}_{nm} \mathcal{A}_m = -\F_n/\s\,,
\label{eq:MAF}
\end{equation}
where
\begin{equation}
   \mathbb{M}_{nm}= \left\{
   \begin{array}{lr}
     \l(\mathcal{K}_0 + \tfrac12\C^{-1} \F_0\r)\delta_{0m} + \tfrac12\C^{-1}
\F_m, \quad &n = 0; \\[1.25em]
     \l(\mathcal{K}_n + \tfrac12\C^{-1} \F_0\r)\delta_{nm} +
\tfrac12\C^{-1}(\F_{m+n} + \F_{m-n}), \quad &n > 0.
   \end{array}
    \right.
\end{equation}
Recall that the Fourier coefficients of $\fl(\theta)$,~$\F_{m}$, are given by~\eqref{eq:flcoeff}.

Clearly, the  the matrix equation~\eqref{eq:MAF} must be inverted numerically for the~$\mathcal{A}_m$. Once we obtain the~$\mathcal{A}_m$ we can recover the Laplace transform of the swimming velocity from
\begin{equation}
  \widetilde\Uc(\s) =
  \frac{\mu}{2\pi}
  \int_{-\pi}^\pi \pd_\theta\tilde\varphi(\RR,\theta,\s)\sin\theta\dint\theta
  = \tfrac12\mu \mathcal{A}_1.
\end{equation}

\begin{figure}
  \includegraphics[width=0.75\textwidth]{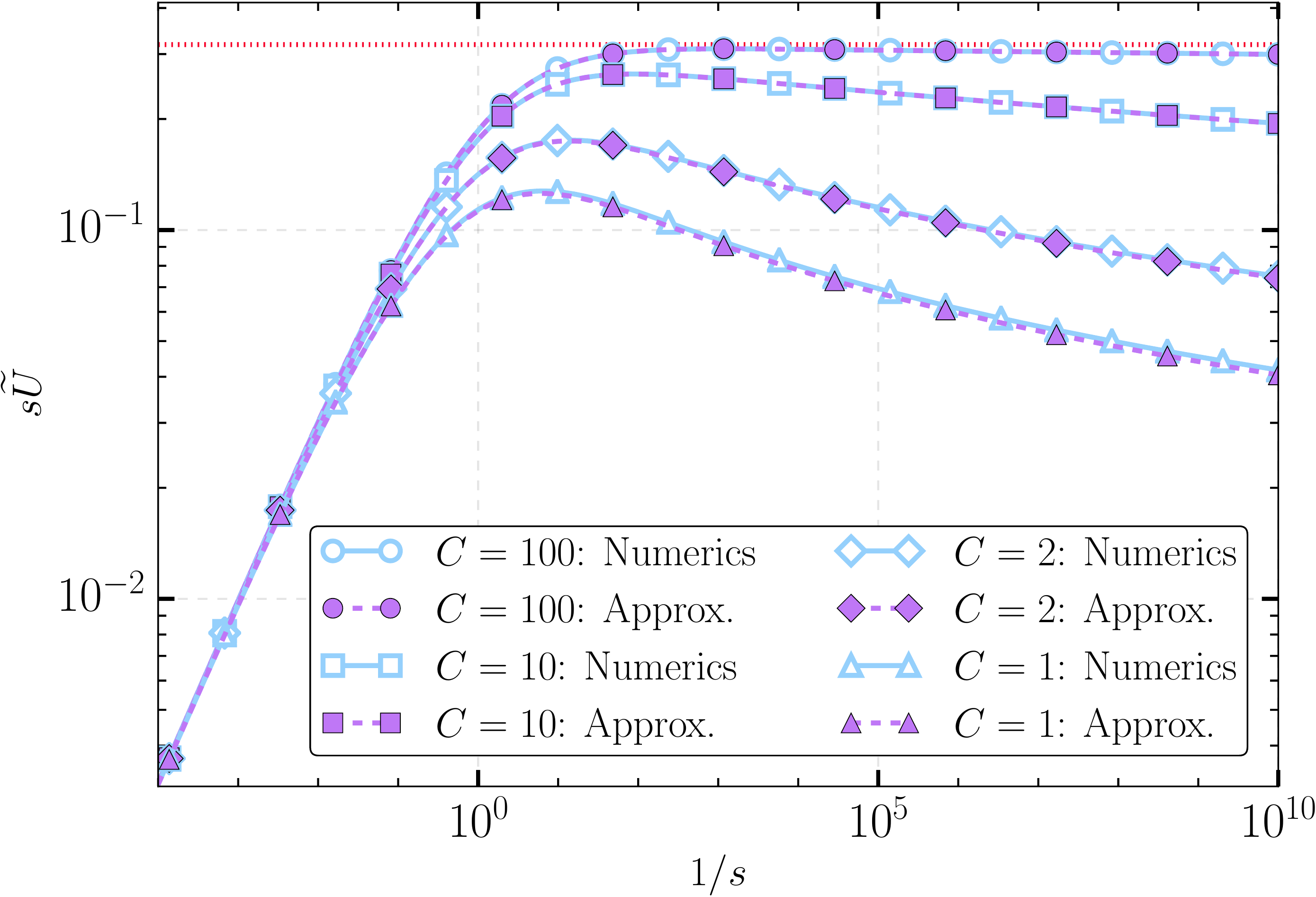}
  \caption{Laplace transform of the swimming velocity as a function of~$1/\s$.
  The limit $1/\s\rightarrow\infty$ of $\s\widetilde\Uc(\s)$ shows the limit of
  $\Uc(\time)$ as~$\time\rightarrow\infty$.
      We use the step function form~\eqref{eq:surf_flux} for~$\fl(\theta)$ and set
    $\RR=\Diff=-\F=\mu=1$.  Empty symbols show the numerical solution to \eqref{eq:MAF} while filled symbols show the approximate solution using only the first two modes $\mathcal{A}_0$ and~$\mathcal{A}_1$.
    The  horizontal red dotted line is
    the~$\C\rightarrow\infty$ time-asymptotic swimming speed,
    Eq.~\eqref{eq:swimvel2D}.}
  \label{fig:U_mixedbc}
\end{figure}

We solve the problem numerically  in the case of the same Janus particle as in Section \ref{sec:BC1} by truncating \eqref{eq:MAF}. Recalling that the constant asymptote of~$\Uc(\time)$ as~$\time\rightarrow\infty$ is~$\s\widetilde\Uc(\s)$ as~$\s\rightarrow0$, we plot $\s\widetilde\Uc(\s)$ as a function of $1/s$  in Figure~\ref{fig:U_mixedbc} for $C$  ranging from 1 to 100 (empty symbols).  For large background concentration~$\C$ (upper curves) the particle does indeed keep a constant swimming velocity for a while, as predicted by Eq.~\eqref{eq:swimvel2D} (horizontal red dotted line).  For lower values of $\C$, the reactant is depleted more quickly, and eventually the swimming speed starts to decrease.  The crossover times~$\T$ are consistent with equation~\eqref{eq:T}: the~$\C=1$ curve in Fig.~\ref{fig:U_mixedbc} has~$\log_{10}\T \sim 0.27$.

We can get a remarkably accurate approximation for~$\widetilde\Uc(\s)$ by retaining only the first two modes, $\mathcal{A}_0$ and~$\mathcal{A}_1$, in~\eqref{eq:MAF}.  We then just have to solve a two-by-two linear system, and we obtain
\begin{equation}
  \s\widetilde\Uc(\s) \simeq
  -\frac{\mu\,\mathcal{K}_0\F_1}
  {\C^{-2}(\F_1^2 - \F_0(2\F_0+\F_2))
    - \C^{-1}(\mathcal{K}_0\F_2 + 2(\mathcal{K}_0 + \mathcal{K}_1)\F_0)
    - 2\mathcal{K}_0\mathcal{K}_1
  }\cdot
  \label{eq:sUth}
\end{equation}
This approximate solution is also plotted in Fig.~\ref{fig:U_mixedbc} (filled symbols); it coincides almost exactly with the numerical solution over the full range of~$\s$.    For~$\C\rightarrow\infty$, most terms vanish in the denominator of~\eqref{eq:sUth}, and we recover~$\s\widetilde\Uc(\s) \sim -\tfrac12\mu\F_1$ after using~$\mathcal{K}_1 \sim -1$.

The small-$\s$ form of the~$\mathcal{K}_m$ is
\begin{equation}
  \mathcal{K}_m \sim \left\{
  \begin{array}{rl}
    1/G(\s^{-1}) + \Order{\s\log\s}, \quad &m = 0; \\[.75em]
    -m + \Order{\s\log\s}, \quad &m>0,
  \end{array}
  \right.
  \label{eq:K012asym}
\end{equation}
with
\begin{equation}
  G(\time) = \tfrac12\log\l(\ee^{2\gamma}\RR^2/4\Diff\time\r).
\end{equation}
From this we derive the long-time asymptotics of~$\Uc(\time)$.
\begin{equation}
  \Uc(\time) \sim
  -\tfrac12\mu\F_1
  \frac{\Cc_1 + 2(\Cc_1\,G(\time) + \Cc_2)}{(\Cc_1\,G(\time) + \Cc_2)^2},
  \qquad
  \time \rightarrow \infty \label{eq:longtimeasymp}
\end{equation}
where
\begin{equation}
  \Cc_1 = \C^{-1}(\C^{-1}(\F_1^2 - \F_0(2\F_0 + \F_2)) + 2\F_0),
\end{equation}
and
\begin{equation}
  \Cc_2 = 2 - \C^{-1}(2\F_0 + \F_2).
\end{equation}
Hence, the swimming velocity goes to zero as~$1/\log\time$.

\section{Phoretic swimming in  two dimensions:  Full problem}
\label{sec:full}
We have so far formulated the problem for determining the swimming speed of a two-dimensional Janus particle by solving an unsteady diffusion equation for the concentration of the reactant.  The swimming speed was then determined by instantaneously integrating the concentration gradient around the surface of the particle.  We have, to this point, neglected however the effects of advection (small P\'eclet number limit).  It is, however, an approximation and we now consider in this section the potential role played by advection of the reactant by the phoretic flow. In Section~\ref{sec:vel}, we solve for the fluid velocity field~$\uv(\rv,\time)$ corresponding to a concentration~$\varphi$ assumed to be known.  The boundary condition on the velocity field is taken to be a known function that represents the concentration at that boundary.  Following this analysis, in Section~\ref{sec:adv} the two problems are coupled together.  That is, we state the equations for a swimming Janus particle with the general boundary conditions formulated in Section~\ref{sec:BC2}, but now the concentration~$\varphi$ and the velocity field are coupled through the boundary conditions for the flow and the advection of the reactant.  We finally solve the coupled problem in   Section~\ref{sec:results}.

\subsection{Velocity field}
\label{sec:vel}

Consider a two-dimensional swimmer moving with velocity~$-\Uc\xuv$.  This velocity could depend on time, but we formally leave out possible time dependence in this section, since it only enters the problem as a parameter.  We will restore time dependence in the following sections.  In the reference frame comoving  with the center of the swimmer  the fluid appears to be moving past a stationary swimmer with far-field velocity~$\Uc\xuv$.  The swimmer is shaped like a disk of radius~$\RR$ and the swimming is caused, as above, by an imposed velocity at the surface of the swimmer.  Moreover, we shall assume that the velocity field is determined by the Stokes equations in which inertial effects are assumed to be negligible.  This is an excellent approximation for phoretic particles due to their small size \cite{Howse2007,Ebbens2011,ebbens2012}.

 The full problem for the two-dimensional flow was originally solved by~\citet{Blake1971} and we only present the main points here.  The governing equations and boundary conditions are,
\begin{subeqnarray}
    \eta\lapl\uv &=& \grad\pres, \qquad \div\uv = 0; \slabel{eq:Stokes} \\
    \uv(\RR,\theta)\cdot\ruv &=& 0; \slabel{eq:imperm} \\
    \uv(\RR,\theta)\cdot\thetauv &= &\g(\theta); \slabel{eq:vtan} \\
    \lim_{\rr\to\infty}\uv(\rr,\theta) &= &\Uc\xuv; \slabel{eq:Uinf}
  \label{eq:veleqs}
\end{subeqnarray}
where~$\pres$ is the pressure, and~$\eta$ is the dynamic viscosity of the fluid.  The flow the steady incompressible Stokes equation, \eqref{eq:Stokes}.
 The boundary condition~\eqref{eq:imperm} states that there is no flow through the surface of the swimmer, as it should be for an impermeable boundary.  The second boundary condition~\eqref{eq:vtan} is a specified tangential velocity that serves as the propulsion mechanism for the swimmer, where the arbitrary function $\g(\theta)$ is  driven by concentration gradients.

 The system of equations and boundary conditions~\eqref{eq:veleqs} can be reformulated using a streamfunction~$\psi$, writing the velocity field as
\begin{equation}
  \uv = \rr^{-1}\pd_\theta\psi\,\ruv - \pd_\rr\psi\,\thetauv,
  \label{eq:vel_stream}
\end{equation}
which gives the system
\begin{subeqnarray}
  \lapl(\lapl\psi) &=& 0; \slabel{eq:biharm} \\
  \pd_\theta\psi(\RR,\theta) &=& 0; \slabel{eq:stream_imperm} \\
  -\pd_\rr\psi(\RR,\theta) &=& \g(\theta) \slabel{eq:stream_tan}; \\
  \lim_{\rr\to\infty}\psi(\rr,\theta)/\rr &= &\Uc\sin\theta \slabel{eq:psi_ff}.
\end{subeqnarray}
Using separation of variables, applying the boundary conditions, and assuming the boundary velocity is given as a Fourier series by
\begin{equation}
 \g(\theta) = \tilde \g_0
 + \sum_{m=1}^\infty (\g_m\sin m\theta + \tilde \g_m\cos m\theta),
\end{equation}
we find the streamfunction
\begin{equation}
 \psi(\rr,\theta) =
 \tfrac12\RR\,(1 - (\rr/\RR)^2)\sum_{m=1}^\infty
 \g_m\,(\RR/\rr)^m\sin m\theta.
 \label{eq:sfswim}
\end{equation}
The velocity field is then given by~\eqref{eq:vel_stream}.  Here we took the~$\g_m$ as given, but through the phoretic boundary condition, it is of course determined by the concentration~$\varphi$, making the flow/concentration problem fully coupled.

\subsection{Swimming phoretic particle with advection}
\label{sec:adv}

Looking back at the asymptotic solution~\eqref{eq:phi0} for the concentration, we see that the reactant depletion front grows radially outward as~$\sqrt{4\Diff\time}$.  The particle swims at speed~$\Uc$ so, assuming the front starts ahead, the particle catches up to the front when~$\time \sim 4\Diff/\Uc^2 \sim 16\Diff/\mu^2\F_1^2$.
Once the particle catches up to the front, it may encounter unconsumed reactant.  This has so far been neglected, since we did not consider the advection of reactant by the flow.  We should therefore solve the full advection-diffusion problem to determine the final outcome.

The PDEs governing the full system are the Stokes equations and the advection-diffusion equation:
\begin{subequations}
\begin{gather}
 \eta\lapl\uv = \grad\pres, \label{eq:stokes_final}
 \qquad
 \div\uv=0; \\
 \pd_\time\varphi + \uv\cdot\grad\varphi = \Diff\lapl\varphi. \label{eq:phi_final}
\end{gather}
The boundary conditions at~$\rr=\RR$ are
\begin{gather}
 -\RR\ruv\cdot\grad\varphi
 = \fl(\theta)(1 + \varphi/\C),\qquad
 \uv\cdot\ruv = 0, \qquad
 \uv\cdot\thetauv = \mu\,\pd_\theta\varphi, \label{eq:full_bcs}
\end{gather}
while the boundary conditions at~$\rr=\infty$ are
\begin{gather}
 \varphi = 0,\qquad
 \uv = \Uc\xuv.\qquad
\end{gather}
\end{subequations}
In Section~\ref{sec:vel} we derived an expression for~$\uv$ given a function~$\g(\theta,\time) = \mu\,\pd_\theta\varphi{(\RR,\theta,\time)}$.   Hence, we can use the analytic solution~\eqref{eq:sfswim} to eliminate~$\uv$ in~\eqref{eq:phi_final}. Consequently, one only needs to solve {the initial value problem}~\eqref{eq:phi_final} numerically for $\varphi{(\rr,\theta,\time)}$ which we do by specifying a zero initial condition and time-integrating until a steady state has been reached.  Additional details on the numerics are presented in the following subsection.

\subsection{Numerical approach}
\label{sec:numerics}
Equation~\eqref{eq:phi_final} is solved using second-order central differences for spatial discretization and a first-order forward Euler method for time discretization.  The simulation domain is taken to be $\left[0,r_{\mathrm{max}}\right]\times\left[0, 2\pi\right]$.  We use $N_{\rr}$ grid points in the radial direction and $N_{\theta}$ grid points in the azimuthal direction.  The flux boundary condition at the surface is implemented via a ghost point to preserve the second-order spatial accuracy of the method.

The analytical expression for the streamfunction~\eqref{eq:sfswim} together with equation~\eqref{eq:vel_stream} is used to determine the velocity field at each time step.  We perform this calculation in Fourier space.  Let $\widehat{(\cdot)}_{m}$ denote the Fourier transform in the azimuthal direction corresponding to mode $m$.  Taking the Fourier transform of~\eqref{eq:uvtheta} for the surface swimming velocity yields
\begin{equation}
  \gh_{m} = \imi\mu m \widehat{\varphi}_{m}.
\end{equation}
The streamfunction~\eqref{eq:sfswim} in Fourier space is
\begin{equation}
  \widehat{\psi}_{m}\left(\rr\right) = \tfrac12\RR(1 - (\rr/\RR)^2)(\RR/\rr)^{m}\,\gh_{m},
\end{equation}
and the velocity components are
\begin{equation}
  \widehat{u}_{m,r}(\rr) = \imi\rr^{-1}m\widehat{\psi}_{m}(\rr), \quad \widehat{u}_{m,\theta}(\rr) = -\pd_\rr\widehat{\psi}_{m}(\rr) \label{eq:fourier_uv}.
\end{equation}
The inverse Fourier transform of the velocity components~\eqref{eq:fourier_uv} provides an exact solution for the velocity field based upon our previously derived analytical solution~\eqref{eq:sfswim}.

All of our simulations consider a Janus particle half-coated with reactant.  To avoid the sharp transitions required by~\eqref{eq:surf_flux} in the numerics, we modify~\eqref{eq:surf_flux} to be a combination of hyperbolic tangent functions.  Therefore, our numerical surface flux boundary condition takes the modified form,
\begin{equation}
  \fl(\theta) = \tfrac12\F\left[\tanh\left(\frac{\theta - 3\pi/2}{\delta}\right) - \tanh\left(\frac{\theta - \pi/2}{\delta}\right)\right] + \F, \quad 0 \leq \theta \leq 2\pi,
\end{equation}
with $\delta = 0.1$.  The derivative~$\fl'(\theta)$ is not periodic, but for small~$\delta$ the jump from~$\theta=2\pi$ to~$0$ is negligible (namely $10^{-13}$ for~$\delta=0.1$).

\begin{figure}
  \centering
  \begin{subfigure}{0.4\textwidth}
    \centering
    \includegraphics[width=\textwidth]{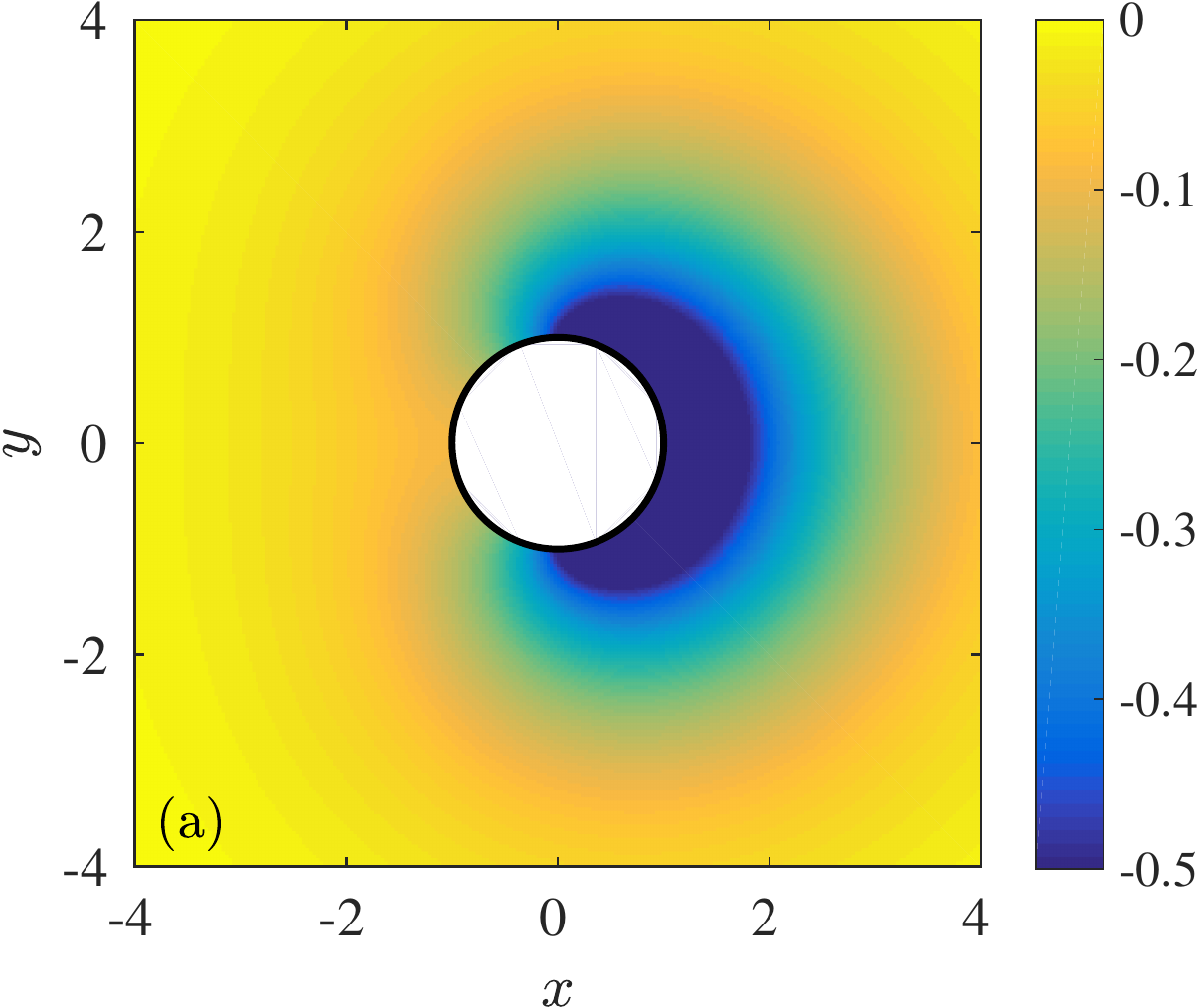}
    \label{fig:phi_no_adv}
  \end{subfigure}
  ~~
  \begin{subfigure}{0.4\textwidth}
    \centering
    \includegraphics[width=\textwidth]{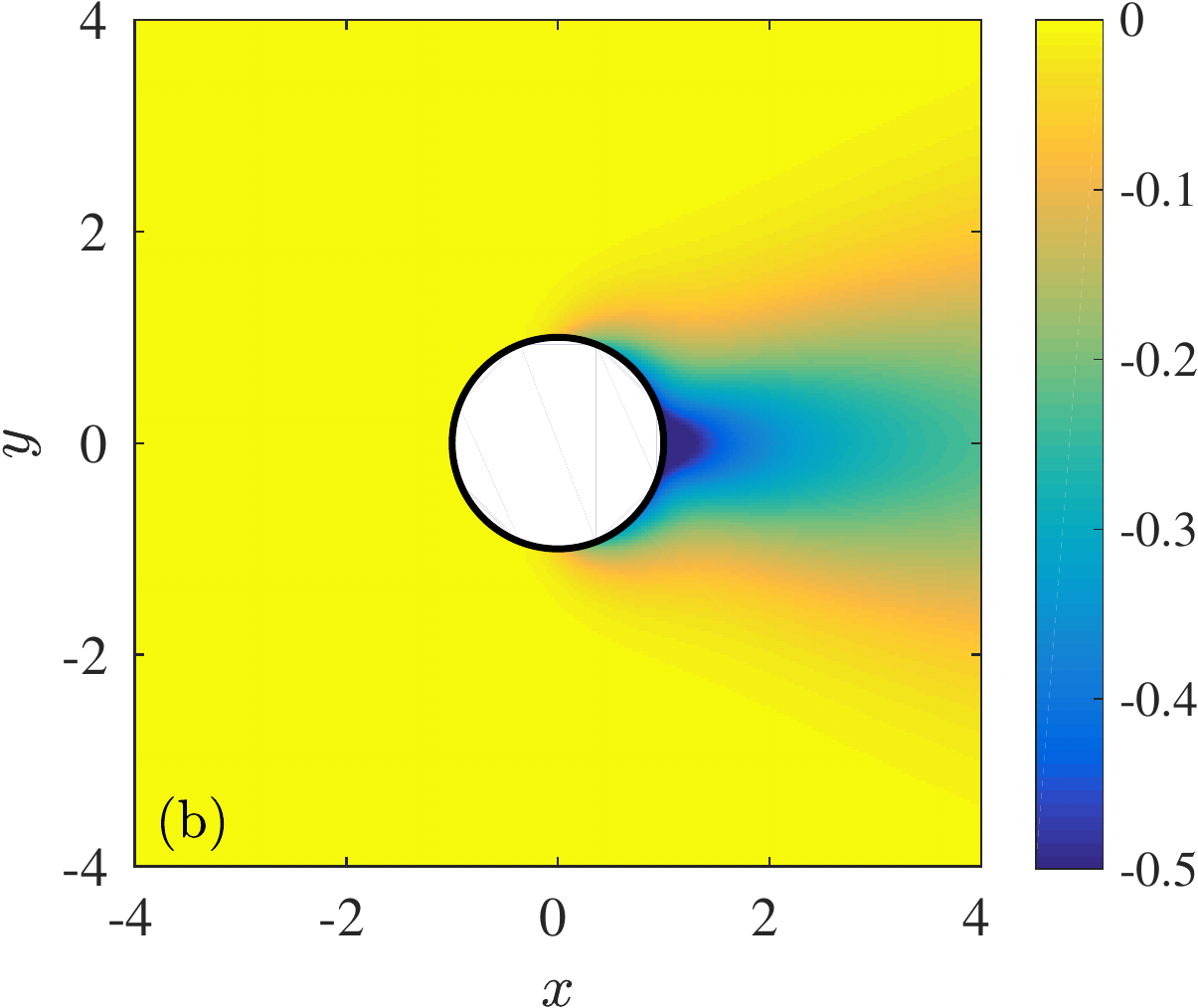}
    \label{fig:phi_adv}
  \end{subfigure}
  \caption{Contour plots of the concentration $\varphi$ (a) without advection and (b) with advection, at $\mu=25$ and $\C=10$.  The advective case has $\Pe=3.45$.}
  \label{fig:phi_contour}
\end{figure}

We use $N_{\theta} = 70$ grid points in the azimuthal direction unless stated otherwise.  The final time, $t_{\mathrm{f}}$ was chosen to be to be sufficiently long to determine a steady swimming velocity.  The radial boundary was set based on the diffusion front so that $r_{\mathrm{max}} = 2\sqrt{4Dt_{\mathrm{f}}}$.  Finally, the time step was chosen to provide sufficient temporal resolution while respecting the CFL condition.  Most of our simulations used $\dt = 2.5\times 10^{-4}$ which yielded a CFL number well below the limit for stability.  We found, however, that such a small time step was necessary to resolve the relevant dynamics of the system such as the overshoot in the swimming velocity shown seen below in Figs.~\ref{fig:UofC} and~\ref{fig:Uofmu}.  In order to validate our code, we compared swimming velocities predicted from our code to the asymptotic result derived earlier in this paper, and found excellent agreement in the appropriate parameter regime.

\subsection{Results with advection}
\label{sec:results}

We now present results pertaining to swimming velocities of two-dimensional Janus particles in the presence of advection.  Figure~\ref{fig:phi_contour} illustrates contour plots of the concentration $\varphi$ without and with advection (left and right plots, respectively) in the case where $\mu=25$ and $\C=10$.  When advection is included, the particle moves into reactant-rich regions which helps it maintain a steady swimming velocity.  The advective case has $\Pe=3.45$.

\begin{figure}[t]
  \includegraphics[width=0.65\textwidth]{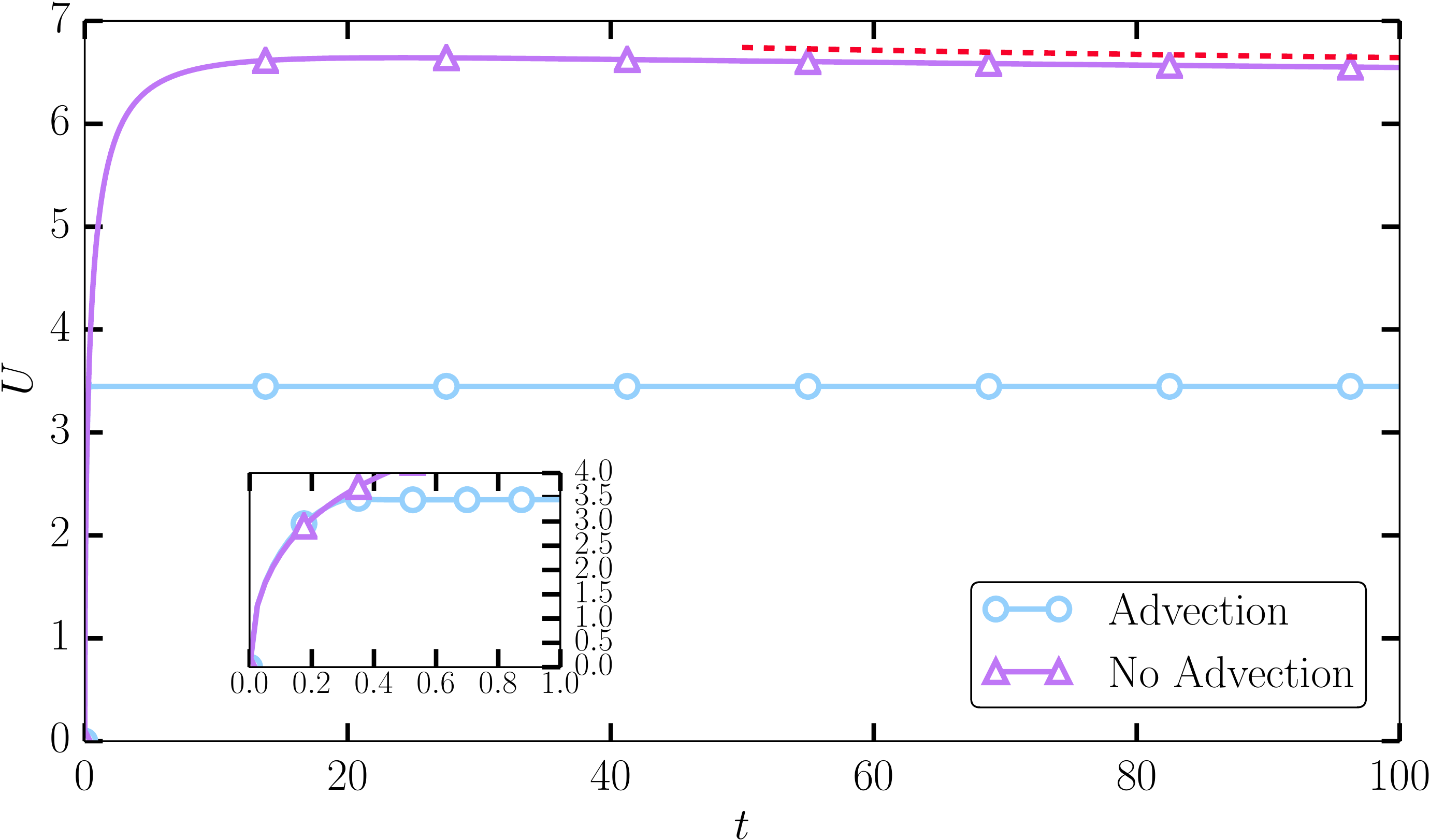}
  \caption{Time evolution of the swimming velocity in a system with (circles) and without advection (triangles) for $\mu=25$ and $\C=10$.
           The dashed line indicates the asymptotically decaying swimming velocity, $\Uc \sim 1/\log\time, \ \ \time\to \infty$ derived in the present work.
           The inset shows a smooth transition to the final swimming velocity at early times.}
  \label{fig:Ucomp}
\end{figure}

 We then compare in Figure~\ref{fig:Ucomp}  the swimming velocity of the Janus particle with and without advection for $\mu = 25$ and $\C = 10$.
Without  advection, the  swimming velocity of the particle gradually decreases to zero according to the asymptotic law derived in Section~\ref{sec:BC2}
as $U\sim 1/\log t$. This fact is emphasized in Figure~\ref{fig:Ucomp} by comparing the analytical result~\eqref{eq:longtimeasymp} (red dashed line) to the numerical solution without advection (triangles).  In stark contrast, when advection is taken into account the phoretic  particle swims with a constant velocity.
The inset in Figure~\ref{fig:Ucomp} shows the smooth transition from zero initial swimming velocity to the final steady swimming velocity.

\begin{figure}[t]
  \includegraphics[width=0.6\textwidth]{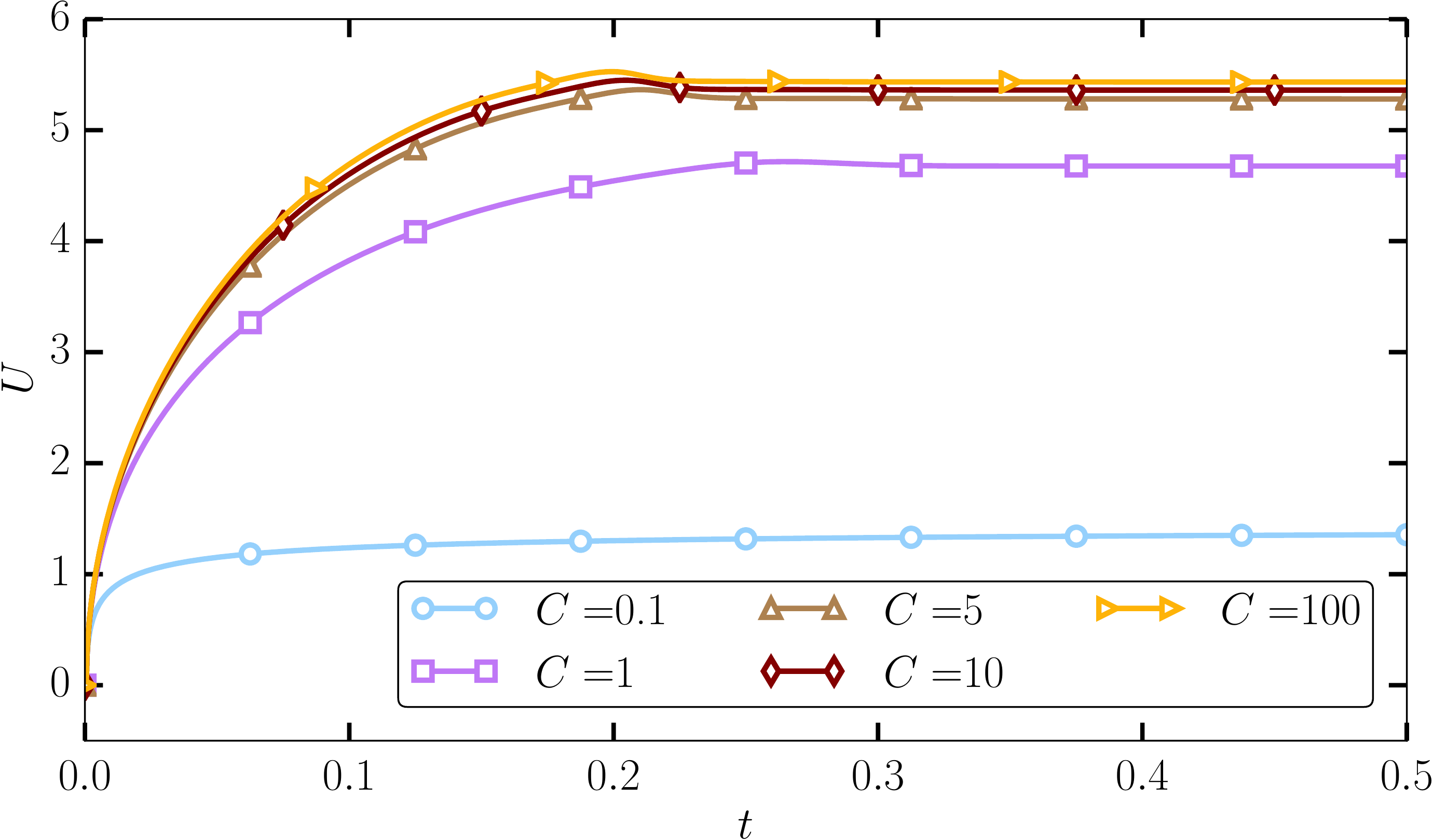}
  \caption{Time evolution of the swimming velocity for various values of the $C$ and $\mu = 50$.}
  \label{fig:UofC}
\end{figure}

The value of this steady swimming velocity depends on~$\mu$ and the concentration of the reactant, $\C$.  We show in Figure~\ref{fig:UofC}  the time evolution of the velocity at $\mu = 50$ for various values of $\C$.  In each case, the Janus particle eventually reaches a constant swimming velocity.  We note that for larger values of $\C$ the swimming velocity exhibits a small overshoot above the final swimming velocity ranging from $0.8\%$ ($\C = 1$) to $1.7\%$ ($\C = 100$) above the final swimming velocity.  This overshoot vanishes as $\C$ is decreased and is no longer present at $\C=0.1$.  We further note that the final swimming velocity appears to become independent of $\C$ as $\C$ becomes large.  This is to be expected since in the limit of infinite reactant the particle has no reason to slow down.

\begin{figure}[t]
  \includegraphics[width=0.6\textwidth]{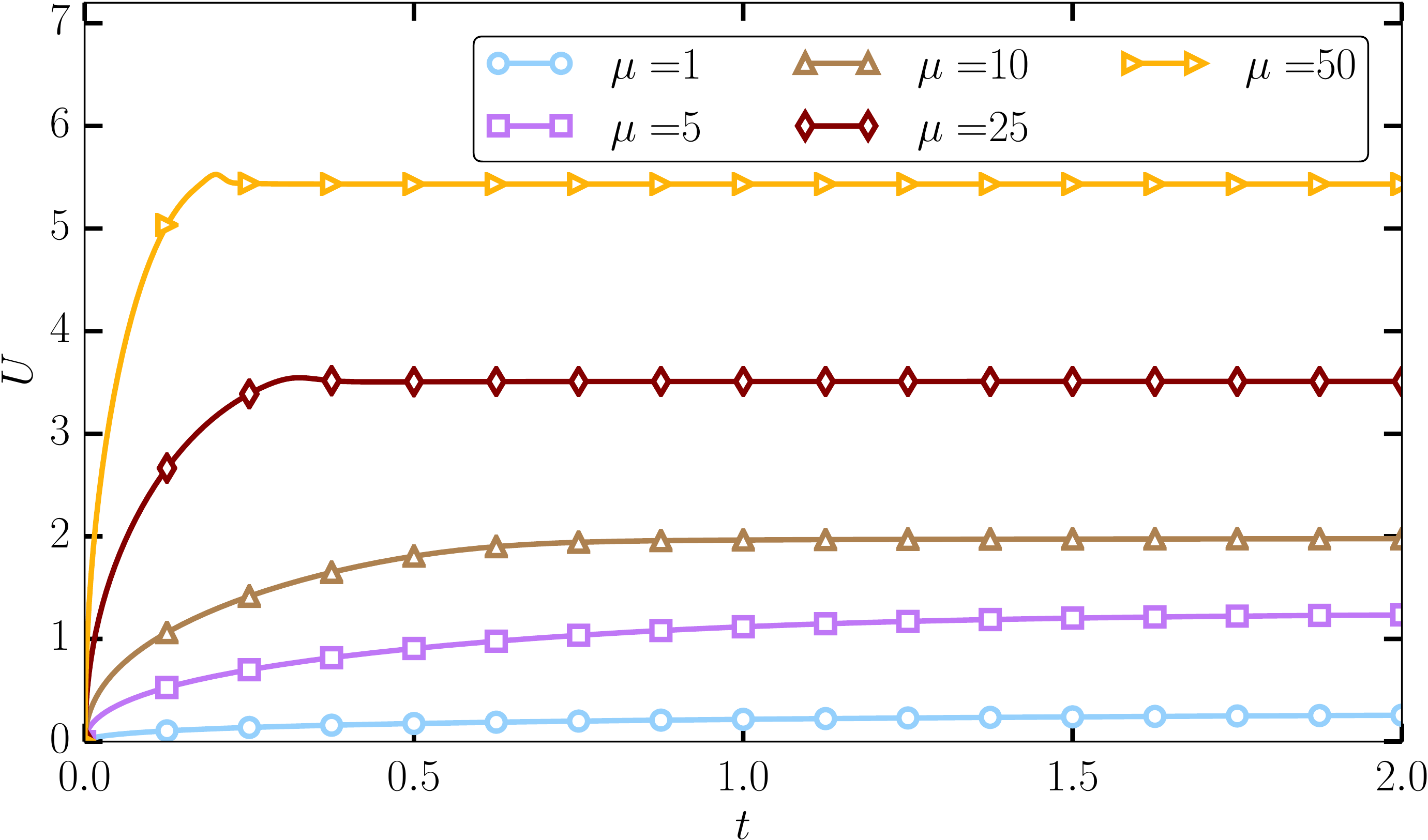}
  \caption{Time evolution of the swimming velocity at various values of $\mu$ for $\C = 100$. }
  \label{fig:Uofmu}
\end{figure}

We next consider the dependence of the swimming velocity on $\mu$ and plot the results in the case where  $\C = 100$ in Fig.~\ref{fig:Uofmu}.  The particle takes  longer to reach its final swimming velocity as $\mu$ is decreased.  Once again, we observe an overshoot in the swimming velocity (ranging from $1\%$ to $1.7\%$ at $\mu = 25$ and $\mu = 50$, respectively).    Finally, we observe that the swimming velocity does not become independent of $\mu$ but increases with it.

\begin{figure}[ht]
  \includegraphics[width=0.6\textwidth]{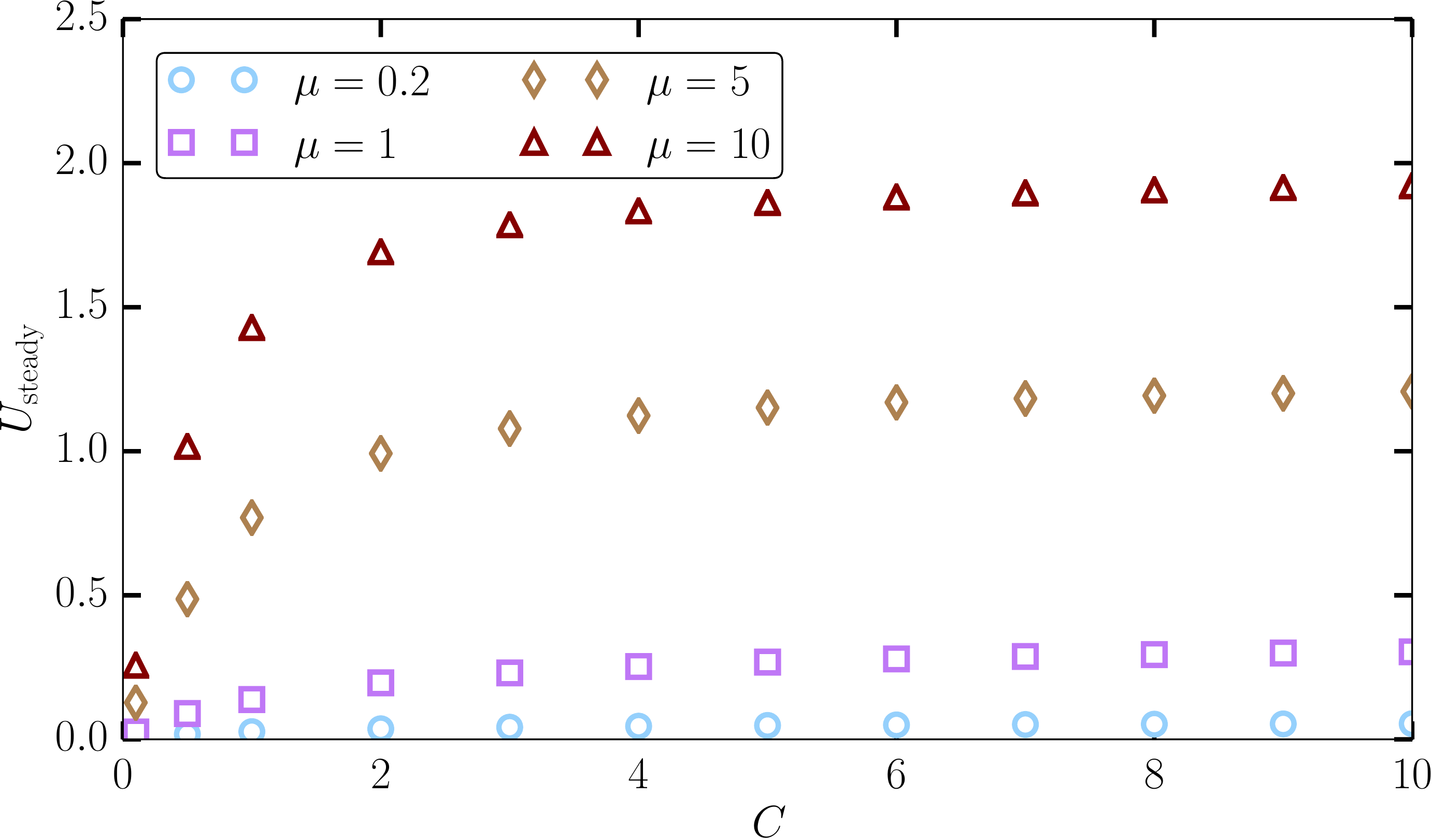}
  \caption{Steady swimming velocity as a function of reactant concentration for several values of $\mu$.  The swimming velocity saturates for large values of $\C$.  For small $\C$ the swimming velocities are linear in $\mu$ whereas they become nonlinear in $\mu$ for large $\C$.}
  \label{fig:Us}
\end{figure}

In Figure~\ref{fig:Us} we finally plot the steady swimming velocity, $\Uc_{\mathrm{steady}}$, as a function of reactant concentration, $\C$, at various values of $\mu$.  In our dimensionless units, we recall that $\Uc_{\mathrm{steady}}$ is the  same as the P\'eclet number $\Pe_{\mathrm{steady}} = \Uc_{\mathrm{steady}}\RR/\Diff$.  For small $\mu$, the steady swimming velocity is very small in accordance with the low-$\Pe$ regime.
  Moreover, $\Uc_{\mathrm{steady}}$ is linear in $\mu$ for small $\C$ (e.g. compare $\Uc_{\mathrm{steady}}$ at $\mu = 5$ and $\mu = 10$ for $\C < 1$) but becomes nonlinear in $\mu$ for large $\C$.  In all cases, $\Uc_{\mathrm{steady}}$ saturates for large $\C$.

\section{Conclusions}
\label{sec:conclusions}

In the present work, we studied the ability of two-dimensional phoretic (Janus) particles to reach a  steady swimming velocity.  Our problem was formulated in terms of a two-dimensional diffusion  equation for the concentration of a reactant around a circular  particle.  The motion  of the particle was driven by reactions at the surface of the particle which were imposed  through a flux boundary condition for the concentration field.  Only one half of the particle was coated with a reactant, with the flux of reactant given by $\F<0$.

Our first general analytical  approach was to perform an asymptotic analysis via Laplace transforms to assess the final swimming speed of the particle when advection of the reactant by the flow was neglected.  A key step in our analysis was use of a Tauberian theorem which provided the asymptotic antiderivative of the Laplace-transformed solution.  We considered two boundary conditions in our analysis which led to two different results.  In our initial approach, we ignored the fact that the particle is immersed in a field of finite reactant concentration.  In that case, the Janus particle reached a steady swimming velocity of $\Uc = -\mu \F / \pi$, as long as it did not exhaust the reactant.  We then generalized the boundary conditions to include the effects of a finite reactant concentration.  With the generalized boundary conditions, our analysis revealed that the Janus particle has an asymptotically-decaying swimming velocity.  In particular, we found that the  swimming  velocity of the particle will decay to zero as $\Uc \sim 1/\log\time$.  The fact that the particle eventually stops swimming is to be expected given that it eventually runs out of reactant to consume.

Next,  we generalized the study to include advective effects due to  motion of the particle.  In this situation, the concentration field around the Janus particle was advected by a velocity field given by Stokes flow.  We analytically solved the Stokes equation for the velocity field.   We then solved the concentration field numerically using the analytically determined velocity field  and the concentration-limited boundary conditions.  In the advective case, we found that the Janus  particle reaches a steady swimming velocity which is linked to the fact that the Janus particle is  continually moving into reactant-rich areas so that it never locally depletes the reactant.  The lack of a steady-state solution for a two-dimensional phoretic particle is thus essentially a result of neglecting chemical advection.

In the future it would be interesting  to analyze different configurations for the chemical coating beyond the simple step function used here (Janus particles).  Different particle geometries, such as ellipses and crescents,  may provide additional insight. In particular, and in analogy with problems in electrostatics,
particles with singularities in their shapes, such as kinks and cusps, could display  interesting amplifications of local chemical gradients.  Another  direction likely to be of interest  would be to model weak three-dimensional effects  associated with particles in a thin film.

\begin{acknowledgments}
  The research was supported by NSF Grants DMS-1109315 and DMS-1147523
  (Madison) and by the European Union through a CIG grant (Cambridge).
\end{acknowledgments}

\bibliography{janus2d}

\end{document}